# ROTATING DUST SOLUTIONS OF EINSTEIN'S EQUATIONS WITH 3-DIMENSIONAL SYMMETRY GROUPS PART 2: ONE KILLING FIELD SPANNED ON $u^\alpha$ AND $w^\alpha$

Andrzej Krasiński

N. Copernicus Astronomical Center and College of Science

Polish Academy of Sciences, Bartycka 18, 00 716 Warszawa, Poland

email: akr@alfa.camk.edu.pl

**Abstract.** This is the second part of a series of 3 papers. Using the same method and the same coordinates as in part 1, rotating dust solutions of Einstein's equations are investigated that possess 3-dimensional symmetry groups, under the assumption that only one of the Killing fields is spanned on the fields of velocity $u^\alpha$ and rotation $w^\alpha$, while the other two define vectors that are linearly independent of $u^\alpha$ and $w^\alpha$ at every point of the spacetime region under consideration. The Killing fields are found and the Killing equations solved for the components of the metric tensor in every case that arises. The Einstein equations are simplified in a few cases, three (most probably) new solutions are found, and several classes of solutions known earlier are identified in the present scheme. They include those by Ozsváth, Maitra, Ellis, King and Vishveshwara and Winicour.

# ROTATING DUST SOLUTIONS OF EINSTEIN'S EQUATIONS WITH 3-DIMENSIONAL SYMMETRY GROUPS PART 2: ONE KILLING FIELD SPANNED ON $u^\alpha$ AND $w^\alpha$

Andrzej Krasiński

N. Copernicus Astronomical Center and College of Science

Polish Academy of Sciences, Bartycka 18, 00 716 Warszawa, Poland

email: akr@alfa.camk.edu.pl

**Abstract.** This is the second part of a series of 3 papers. Using the same method and the same coordinates as in part 1, rotating dust solutions of Einstein's equations are investigated that possess 3-dimensional symmetry groups, under the assumption that only one of the Killing fields is spanned on the fields of velocity $u^\alpha$ and rotation $w^\alpha$, while the other two define vectors that are linearly independent of $u^\alpha$ and $w^\alpha$ at every point of the spacetime region under consideration. The Killing fields are found and the Killing equations solved for the components of the metric tensor in every case that arises. The Einstein equations are simplified in a few cases, three (most probably) new solutions are found, and several classes of solutions known earlier are identified in the present scheme. They include those by Ozsváth, Maitra, Ellis, King and Vishveshwara and Winicour.

**I. Summary of the method.**

This is a concise summary of results that will be used in this paper. For proofs, motivations and references see Paper 1 [1].

Every timelike vector field $u^\alpha$ of unit length that has zero acceleration and nonzero rotation defines the functions $\tau(x), \eta(x)$ and $\xi(x)$ such that:

$$u_\alpha = \tau_{,\alpha} + \eta \xi_{,\alpha} . \tag{1.1}$$

These functions are defined up to the transformations:

$$\tau = \tau' - S(\xi', \eta'), \qquad \xi = F(\xi', \eta'), \qquad \eta = G(\xi', \eta'), \tag{1.2}$$

where the functions $F$ and $G$ obey:

$$F_{,\xi'} G_{,\eta'} - F_{,\eta'} G_{,\xi'} = 1, \tag{1.3}$$

(this guarantees that the Jacobian of the transformation is 1), and $S$ is determined by:

$$S_{,\xi'} = GF_{,\xi'} - \eta', \qquad S_{,\eta'} = GF_{,\eta'} . \tag{1.4}$$

If $u^\alpha$ is the velocity field of a fluid whose number of particles is conserved:

$$(\sqrt{-g} n u^\alpha)_{,\alpha} = 0, \tag{1.5}$$

(where $g$ is the determinant of the metric tensor and $n$ is the particle number density), then one more function $\zeta(x)$ exists such that:

$$\sqrt{-g} n u^\alpha = \varepsilon^{\alpha\beta\gamma\delta} \xi_{,\beta} \eta_{,\gamma} \zeta_{,\delta} , \tag{1.6}$$

and it is determined up to the transformations:

$$\zeta = \zeta' + T(\xi', \eta'). \tag{1.7}$$

Note that $n$ is not defined uniquely by (1.5). For example, if $u^\alpha = \delta^\alpha{}_0$ and $n$ obeys (1.5), then $n' = nf(x,y,z)$ (where $f$ is an arbitrary function) will also obey (1.5). This nonuniqueness allows for a greater freedom in the choice of $\zeta$ than (1.7), and the freedom will be used in some cases.

The following relations hold:

$$u^\alpha \tau_{,\alpha} = 1, \qquad u^\beta \xi_{,\beta} = u^\beta \eta_{,\beta} = u^\beta \zeta_{,\beta} = 0.$$

$$\frac{\partial(\tau, \eta, \xi, \zeta)}{\partial(x^0, x^1, x^2, x^3)} = \sqrt{-g}n \neq 0. \tag{1.8}$$

The last of (1.8) guarantees that $\{\tau, \xi, \eta, \zeta\}$ can be chosen as coordinates, they will be called Plebański coordinates. Then, with $\{\tau, \xi, \eta, \zeta\} = \{x^0, x^1, x^2, x^3\} = \{t, x, y, z\}$:

$$u^\alpha = \delta^\alpha{}_0, \qquad u_\alpha = \delta^0{}_\alpha + y\delta^1{}_\alpha,$$

$$g_{00} = 1, \qquad g_{01} = y, \qquad g_{02} = g_{03} = 0, \qquad g = \det(g_{\alpha\beta}) = -n^{-2},$$

$$w^\alpha = n\delta^\alpha_3, \qquad \omega_{\alpha\beta} = -\omega_{\beta\alpha} = (1/2)\delta^1{}_\alpha \delta^2{}_\beta, \tag{1.9}$$

where $w^\alpha$ is the rotation vector field, and $\omega_{\alpha\beta}$ is the rotation tensor corresponding to the velocity field $u_\alpha$:

$$\omega_{\alpha\beta} = \frac{1}{2}(u_{\alpha,\beta} - u_{\beta,\alpha} - \dot{u}_\alpha u_\beta + \dot{u}_\beta u_\alpha), \qquad w^\alpha = -(1/\sqrt{-g})\varepsilon^{\alpha\beta\gamma\delta}u_\beta \omega_{\gamma\delta}. \tag{1.10}$$

If $\omega_{\alpha\beta} \neq 0$ and $\dot{u}^\alpha = 0$ (what is assumed throughout), then necessarily the pressure $p =$ const and $\kappa p$ may be interpreted as the cosmological constant ($\kappa := 8\pi G/c^4$).

If any Killing vector field exists on a manifold (on which all the assumptions specified so far are fulfilled), then, in the coordinates of (1.9), it must be of the form:

$$k^\alpha = (C + \phi - y\phi_{,y})\delta^\alpha{}_0 + \phi_{,y}\delta^\alpha{}_1 - \phi_{,x}\delta^\alpha{}_2 + \lambda\delta^\alpha{}_3, \tag{1.11}$$

where $C$ is an arbitrary constant and $\phi(x,y)$ and $\lambda(x,y)$ are arbitrary functions of two coordinates. Whenever $\phi_{,\alpha} \neq 0$, a transformation of the class (1.2) - (1.4) can be found that leads to:

$$k^\alpha = \delta^\alpha{}_1. \tag{1.12}$$

The metric then becomes independent of $x$, and the coordinates preserving (1.12) are determined up to the transformations:

$$t' = t - \int yH_{,y}\,dy + A, \qquad x' = x + H(y), \qquad y' = y, \qquad z' = z + T(y), \tag{1.13}$$

where $A$ is an arbitrary constant and $H, T$ are arbitrary functions.

The condition $\phi_{,\alpha} \neq 0$ that allows one to fulfil (1.12) means that the Killing vector $k^\alpha$ is linearly independent of the vectors $u^\alpha$ and $w^\alpha$ at every point of the spacetime region under consideration. In Paper 1, solutions of the Killing equations and of the Einstein equations were considered under the assumption that there exist three Killing vector fields on the manifold, two of which have $\phi = $ const in (1.11), while the third one has $\phi_{,\alpha} \neq 0$ and can be transformed to the form (1.12). In the present paper it is assumed that only one Killing field has $\phi = $ const (it will be $k_{(3)}$, see below), and two have $\phi_{,\alpha} \neq 0$. Only one of the two can then be transformed to the form (1.12) (it will be $k_{(1)}$), the other one ($k_{(2)}$) will preserve its general form (1.11).

In the whole paper, whenever reference is made to the Einstein tensor, the components quoted are projections of the coordinate components $G_{\alpha\beta}$ onto the orthonormal tetrad $e_i{}^\alpha$, i.e. $G_{ij} = e_i{}^\alpha e_j{}^\beta G_{\alpha\beta}$. In every case it will be self-evident how the tetrad is defined.

## II. The Lie algebra of the symmetry group.

According to the assumptions made in the preceding section, there exist the following three Killing vector fields:

$$k_{(1)}^\alpha = \delta_1^\alpha,$$
$$k_{(2)}^\alpha = (C_2 + \phi - y\phi_{,y})\delta_0^\alpha + \phi_{,y}\delta_1^\alpha - \phi_{,x}\delta_2^\alpha + \lambda_2(x,y)\delta_3^\alpha,$$
$$k_{(3)}^\alpha = C_3\delta_0^\alpha + \lambda_3(x,y)\delta_3^\alpha, \tag{2.1}$$

where $C_2$ and $C_3$ are arbitrary constants, and $\phi$, $\lambda_2$ and $\lambda_3$ are unknown functions of $(x,y)$, to be determined from the commutation relations. The coordinates of (2.1) are determined up to (1.13).

The fields $k_{(1)}, k_{(2)}$ and $k_{(3)}$ will form a Lie algebra if constants $a,\ldots,j$ exist such that:

$$[k_{(1)}, k_{(2)}] = ak_{(1)} + bk_{(2)} + ck_{(3)},$$
$$[k_{(1)}, k_{(3)}] = dk_{(1)} + ek_{(2)} + fk_{(3)},$$
$$[k_{(2)}, k_{(3)}] = gk_{(1)} + hk_{(2)} + jk_{(3)}, \tag{2.2}$$

Eqs. (2.2) are equivalent to the following set:

$$\phi_{,x} - y\phi_{,xy} = b(C_2 + \phi - y\phi_{,y}) + cC_3, \tag{2.3a}$$

$$\phi_{,xy} = a + b\phi_{,y}, \tag{2.3b}$$

$$\phi_{,xx} = b\phi_{,x}, \tag{2.3c}$$

$$\lambda_{2,x} = b\lambda_2 + c\lambda_3, \tag{2.3d}$$

$$e(C_2 + \phi - y\phi_{,y}) + fC_3 = 0, \tag{2.3e}$$

$$d + e\phi_{,y} = 0, \tag{2.3f}$$

$$e\phi_{,x} = 0, \tag{2.3g}$$

$$\lambda_{3,x} = e\lambda_2 + f\lambda_3, \tag{2.3h}$$

$$h(C_2 + \phi - y\phi_{,y}) + jC_3 = 0, \tag{2.3i}$$

$$g + h\phi_{,y} = 0, \tag{2.3j}$$

$$h\phi_{,x} = 0, \tag{2.3k}$$

$$\phi_{,y}\lambda_{3,x} - \phi_{,x}\lambda_{3,y} = h\lambda_2 + j\lambda_3. \tag{2.3l}$$

A few of these equations imply alternatives that will have to be considered separately. The alternatives will be organized into a binary tree and numbered in a positional system that will enable one to quickly identify complementary cases (see Fig.1). Of the two alternatives implied by (2.3g) we choose for the beginning the following.

**Case 1:** $\phi_{,x} \neq 0$.
This implies, from eqs. (g), (k), (f) and (j) in (2.3):

$$e = h = d = g = 0. \tag{2.4}$$

Eqs. (2.3a-d) survive unchanged, while the remaining ones simplify to:

$$fC_3 = 0, \tag{2.5e}$$

$$\lambda_{3,x} = f\lambda_3, \tag{2.5h}$$

$$jC_3 = 0, \tag{2.5i}$$

$$\phi_{,y}\lambda_{3,x} - \phi_{,x}\lambda_{3,y} = j\lambda_3, \tag{2.5l}$$

other equations being fulfilled identically. Now we choose the following.

**Case 1.1:** $C_3 \neq 0$,
which implies, from (2.5):

$$f = j = 0, \tag{2.6}$$

$$\lambda_3 = \lambda_3(y), \tag{2.7h}$$

$$\phi_{,x}\lambda_{3,y} = 0. \tag{2.7l}$$

With (2.4) and (2.6), the second and third commutator in (2.2) are zero. Since in Case 1 $\phi_{,x} \neq 0$ by assumption, eqs. (2.7h) and (2.7l) imply:

$$\lambda_3 = \text{const.} \tag{2.8}$$

The solution of (2.3.d) depends on whether $b = 0$ or $b \neq 0$, so we follow now:

**Case 1.1.1:** $b \neq 0$.
Then, from (2.3d):

$$\lambda_2 = \beta(y)e^{bx} - c\lambda_3/b, \tag{2.9}$$

and from (2.3a-c):

$$\phi = \alpha(y)e^{bx} - (a/b)y - C_2 - (c/b)C_3, \tag{2.10}$$

where $\alpha(y)$ and $\beta(y)$ are arbitrary functions. The assumption $\phi_{,x} \neq 0$ implies $\alpha \neq 0$. The basis of the Killing vector fields can be changed to $k_{(1)}$, $k'_{(2)} = k_{(2)} + (c/b)k_{(3)} + (a/b)k_{(1)}$ and $k_{(3)}$. In the new basis:

$$k'^{\alpha}_{(2)} = e^{bx}[(\alpha - y\alpha_{,y})\delta^{\alpha}_0 + \alpha_{,y}\delta^{\alpha}_1 - b\alpha\delta^{\alpha}_2 + \beta\delta^{\alpha}_3] \tag{2.11}$$

(i.e. by taking $k'_{(2)}$ instead of $k_{(2)}$, $c = 0$ was achieved). The transformation (1.13) with $H = (1/b)\ln\alpha$, $T = (1/b)\int(\beta/\alpha)dy$ will now have the same effect as if $\beta = 0$, $\alpha = 1$. Finally, then, the following basis of the Killing vector fields resulted:

$$k^{\alpha}_{(1)} = \delta^{\alpha}_1, \qquad k^{\alpha}_{(2)} = e^{bx}(\delta^{\alpha}_0 - b\delta^{\alpha}_2), \qquad k^{\alpha}_{(3)} = C_3\delta^{\alpha}_0 + \lambda_3\delta^{\alpha}_3, \tag{2.12}$$

but from now on the transformations (1.13) are allowed only with $H$ and $T$ being constants. The commutation relations are:

$$[k_{(1)}, k_{(3)}] = [k_{(2)}, k_{(3)}] = 0, \qquad [k_{(1)}, k_{(2)}] = bk_{(2)}, \tag{2.13}$$

and they correspond to Bianchi type III. In virtue of the commutator $[k_{(1)}, k_{(3)}]$ being zero, coordinates can be adapted to $k_{(1)}$ and $k_{(3)}$ simultaneously so that $k^{\alpha'}_{(1)} = \delta^{\alpha'}_1$, $k^{\alpha'}_{(3)} = \delta^{\alpha'}_0$. The transformation is:

$$(t', x', y') = (t, x, y), \qquad z' = -\lambda_3 t + C_3 z; \tag{2.14}$$

it is nonsingular because Case 1.1 was defined by $C_3 \neq 0$, but it leads out of the Plebański class of coordinates used so far. In the new coordinates (primes dropped), the metric is independent of $x$ and $t$, and:

$$\begin{aligned} k^{\alpha}_{(2)} &= e^{bx}(\delta^{\alpha}_0 - b\delta^{\alpha}_2 - \lambda_3\delta^{\alpha}_3), \\ u^{\alpha} &= \delta^{\alpha}_0 - \lambda_3\delta^{\alpha}_3, \qquad w^{\alpha} = nC_3\delta^{\alpha}_3, \\ g_{00} &= 1 + \lambda_3{}^2 g_{33}, \qquad g_{01} = y + \lambda_3 g_{13}, \\ g_{02} &= \lambda_3 g_{23}, \qquad g_{03} = \lambda_3 g_{33}. \end{aligned} \tag{2.15}$$

Now the Killing equations involving $k_{(2)}$ have to be solved for the components of the metric tensor. For this purpose, two cases have to be considered separately: $\lambda_3 \neq 0$ and $\lambda_3 = 0$.

**Case 1.1.1.1:** $\lambda_3 \neq 0$.

It is convenient to change the coordinates yet again to:

$$(t', x') = (t, x), \qquad Y = \lambda_3 y + bz, \qquad Z = \lambda_3 y - bz. \tag{2.16}$$

The vector fields $u^{\alpha}$, $w^{\alpha}$ and $k^{\alpha}_{(i)}$ become (primes dropped):

$$\begin{aligned} u^{\alpha} &= \delta^{\alpha}_0 - b\lambda_3(\delta^{\alpha}_2 - \delta^{\alpha}_3), \qquad w^{\alpha} = bC_3 n(\delta^{\alpha}_2 - \delta^{\alpha}_3), \\ k^{\alpha}_{(1)} &= \delta^{\alpha}_1, \qquad k^{\alpha}_{(3)} = \delta^{\alpha}_0, \qquad k^{\alpha}_{(2)} = e^{bx}(\delta^{\alpha}_0 - 2b\lambda_3\delta^{\alpha}_2), \end{aligned} \tag{2.17}$$

and the metric tensor, after the transformation (2.16) and after solving the Killing equations becomes (primes dropped again):

$$\begin{aligned} g_{00} &= 1 + (\lambda_3 b)^2(g_{22} + g_{33} - 2g_{23}), \\ g_{01} &= \frac{1}{2\lambda_3}(Y + Z) - \frac{1}{2}b^2\lambda_3 Y(g_{22} - g_{33}) + \lambda_3 h_{13}, \\ g_{02} &= b\lambda_3(g_{22} - g_{23}), \qquad g_{03} = b\lambda_3(g_{23} - g_{33}), \end{aligned}$$

$$g_{11} = \left(\frac{Y}{2\lambda_3}\right)^2 + \frac{1}{2\lambda_3^2}YZ + \left(\frac{1}{2}bY\right)^2(g_{22} + g_{33} + 2g_{23}) - \frac{b}{\lambda_3}h_{12}Y + h_{11},$$

$$g_{12} = -\frac{1}{2}bY(g_{22} + g_{23}) + \frac{1}{2\lambda_3}h_{12} + \frac{1}{2b}h_{13},$$

$$g_{13} = -\frac{1}{2}bY(g_{23} + g_{33}) + \frac{1}{2\lambda_3}h_{12} - \frac{1}{2b}h_{13}, \qquad (2.18)$$

where $h_{11}(Z)$, $h_{12}(Z)$, $h_{13}(Z)$, $g_{22}(Z)$, $g_{23}(Z)$ and $g_{33}(Z)$ are functions to be determined from the Einstein equations.

No progress was made with the Einstein equations in full generality in this case. Progress may be possible with additional simplifying assumptions (e.g. orthogonality relations among the Killing fields).

No solutions belonging to this case have been identified in the existing literature.

**III. Case 1.1.1.2:** $\lambda_3 = 0$.

We go back to (2.15). With $\lambda_3 = 0$, the coordinates of (2.15) are still in the Plebański class (the factor $C_3$ in $w^\alpha$ is allowable because $n$ is not defined uniquely by (1.5), see comment after (1.7)). The formulae (2.15) still apply, with $\lambda_3 = 0$, and the Killing equations involving $k_{(2)}$ are solved by:

$$g_{11} = y^2 + (by)^2 g_{22}(z) - 2by h_{12}(z) + h_{11}(z),$$
$$g_{12} = -byg_{22} + h_{12}, \qquad g_{13} = -byg_{23}(z) + h_{13}(z), \qquad (3.1)$$

the $h_{ij}(z)$, $g_{22}(z)$, $g_{23}(z)$ and $g_{33}(z)$ being arbitrary functions. The following inequalities hold:

$$g_{33} < 0, \qquad g_{22} < 0, \qquad g_{33} - g_{23}^2/g_{22} := h_{33} < 0 \qquad (3.2)$$

for the following reasons, respectively: (i) the rotation field must be spacelike everywhere, (ii) and (iii) the determinant of the metric $g < 0$ or else the metric has an unphysical signature. In virtue of (3.2) one can define the functions $K_{11}, K_{22}, K_{33}$ and $H_{12}, H_{13}, H_{23}$ by:

$$K_{11}^2 = h_{12}^2/g_{22} + (h_{13} - h_{12}g_{23}/g_{22})^2/h_{33} - h_{11},$$
$$K_{22}^2 = -g_{22}, \qquad K_{33}^2 = -h_{33},$$
$$H_{12} = h_{12}/g_{22}, \qquad H_{23} = g_{23}/g_{22},$$
$$H_{13} = h_{13}/h_{33} - h_{12}g_{23}/(g_{22}h_{33}), \qquad (3.3)$$

the appropriate sign of $K_{11}^2$ being guaranteed by the signature $(+---)$ assumed throughout[1]. With these definitions, the metric form (2.15) - (3.1) may be written as:

$$ds^2 = (dt + ydx)^2 - (K_{11}dx)^2 - \{K_{22}[(-by + H_{12})dx + dy + H_{23}dz]\}^2$$
$$- [K_{33}(H_{13}dx + dz)]^2. \qquad (3.4)$$

---

[1] Definitions analogous to (3.3) will be introduced in several other cases in this paper. In each case, the appropriate inequalities are fulfilled for the same reasons.

Eq. (3.4) defines the orthonormal tetrad used in this section, and the tetrad is comoving, i.e. the velocity field coincides with the tetrad vector #0: $u_\alpha = e^0{}_\alpha$. Without loss of generality it may be assumed that:

$$K_{33} = 1, \qquad (3.5)$$

because this result is achieved by the coordinate transformation $z' = \int K_{33} dz$ and redefinitions of $H_{13}$ and $H_{23}$. The transformation leads to a rescaling of $w^\alpha$ in (2.15), but, taking into account the nonuniqueness of the definition of $n$ through (1.5), the new coordinates are effectively still in the Plebański class.

The Einstein equation $G_{01} = 0$ is now integrated with the result:

$$H_{23} = A_{23} K_{11}/K_{22}, \qquad (3.6)$$

($A$ and $B$ with indices will denote arbitrary constants). Eq. (3.6) substituted into $G_{03} = 0$ implies $bA_{23}/(2K_{11}K_{22}) = 0$, and since this is within the case 1.1.1 defined by $b \neq 0$, the result is:

$$A_{23} = 0 = H_{23}. \qquad (3.7)$$

Now $G_{02} = 0$ implies:

$$H_{13} = A_{13} K_{11} K_{22}, \qquad (3.8)$$

and $G_{12} = 0$ implies:

$$H_{12,z} = B_{12} K_{11}/K_{22}{}^3. \qquad (3.9)$$

Using (3.7) - (3.9) in $G_{23} = 0$ one obtains $bB_{12}/(K_{11}K_{22}{}^2) = 0$, and so:

$$B_{12} = 0, \qquad H_{12} = \text{const}. \qquad (3.10)$$

The coordinate transformation $t = t' - H_{12}x/b$, $y = y' + H_{12}/b$ has the same effect as if:

$$H_{12} = 0, \qquad (3.11)$$

which will be assumed from now on[2]. In further analysis two subcases have to be considered separately:

**Case 1.1.1.2.1:** $A_{13} \neq 0$.

It is convenient to go over to the standard Bianchi-type coordinates in which $g_{33} = -1$ while $g_{03} = g_{13} = g_{23} = 0$. The following transformations do it: in the first step we transform:

$$t = T - F(z)Y, \qquad x = X + [\ln(bF)]/b, \qquad y = bYF, \qquad (3.12)$$

where $F(z)$ is determined by:

$$F_{,z}/F = -bA_{13}K_{22}/[K_{11}(1 + A_{13}{}^2 K_{22}{}^2)] \qquad (3.13)$$

---

[2] The transformation preserves $k_{(2)}^\alpha, k_{(3)}^\alpha, u^\alpha, w^\alpha$ and transforms $k_{(1)}^\alpha$ into $k'^\alpha_{(1)} = \delta^\alpha_1 + (H_{12}/b)k^\alpha_{(3)}$, i.e. in the Killing fields $H_{12} = 0$ can be achieved by a change of basis in the Lie algebra.

(the right-hand side is finite and nonzero because of assumptions made earlier). This leads to $g_{0a} = 0$. Then we transform $z = H(Z)$, where:

$$\frac{dH}{dZ} = (1 + A_{13}{}^2 K_{22}{}^2)^{1/2} \tag{3.14}$$

and thereby achieve $g_{33} = -1$. Denoting:

$$L_{11} = K_{11}\frac{dH}{dZ}, \qquad L_{22} = bFK_{22}, \tag{3.15}$$

we obtain the following metric form:

$$ds^2 = [dT + F(Z)(bYdX - dY)]^2 - [L_{11}(Z)dX]^2 - [L_{22}(Z)(-bYdX + dY)]^2 - dZ^2, \tag{3.16}$$

the coordinates no longer being in the Plebański class.

The Einstein equation $G_{02} = 0$ implies:

$$F_{,Z} = AL_{22}/L_{11}, \tag{3.17}$$

where $A$ is an arbitrary constant. Eq. (3.17) can be integrated if the new variable $z'$ is introduced by:

$$z'_{,Z} = L_{22}/L_{11}; \tag{3.18}$$

then:

$$F(z') = Az' + B, \qquad B = \text{const.} \tag{3.19}$$

With $z'$ in place of $Z$, the equation $G_{13} = 0$ is solved by:

$$L_{11} = CL_{22}\exp[-\frac{1}{2}\int (AF/L_{22}{}^2)dz']. \tag{3.20}$$

The only tetrad components of the Einstein tensor that are still nonzero are the diagonal ones. It may be verified that $G_{11} - G_{22} \equiv 0$ in virtue of (3.17) - (3.20), while $G_{11} - G_{33} = 0$ follows by differentiation of $G_{33} = \Lambda$ in virtue of (3.20). Hence, the latter is the last equation to solve, and it is:

$$(L_{22}L_{22,z'})^2 - \frac{1}{2}AFL_{22}L_{22,z'} + (b^2 + A^2/4)L_{22}{}^2 - \frac{1}{4}(bF)^2 = \Lambda(L_{11}L_{22})^2. \tag{3.21}$$

Eq. (3.21) is an integro-differential equation. Substituting for $L_{11}$ from (3.20) and differentiating by $z'$, it can be changed into an ordinary second-order diferential equation that determines $L_{22}(z')$, and then $L_{11}$ can be (in principle) calculated from (3.20). No further progress in solving the Einstein equations was achieved in the general case.

The coordinates of (3.16) are still comoving, while the rotation vector field is:

$$w^\alpha = L_{11}{}^{-2}(-A\delta_1^\alpha - bAY\delta_2^\alpha + bF\delta_3^\alpha). \tag{3.22}$$

The case $A = 0$ that is seen to be simpler is equivalent (via (3.17) and (3.13)) to $A_{13} = 0$ or $b = 0$, and these were set aside for separate consideration. The matter density is:

$$\kappa\epsilon = -3\Lambda + [bF/(L_{11}L_{22})]^2 + (A/L_{11})^2. \tag{3.23}$$

When $\Lambda = 0$, an elementary solution exists. Eq. (3.21) has then the integral:

$$L_{22}{}^2 = \frac{1}{4}[F^2 - (A^2 + 4b^2)(z' + D)^2], \tag{3.24}$$

where $D$ is an arbitrary constant, and $L_{11}$ is found from (3.20) to be:

$$L_{11} = CW_1^{\alpha_1} W_2^{\alpha_2}, \tag{3.25}$$

where:

$$\alpha_{1,2} = \frac{1}{8b^2}[(A^2 + 4b^2)^{1/2} \pm A]^2, \tag{3.26}$$

and $W_1, W_2$ are the two factors of (3.24):

$$W_{1,2} = \frac{1}{2}[F \mp (A^2 + 4b^2)^{1/2}(z' + D)]. \tag{3.27}$$

A (modestly thorough) search through the literature by this author[3] has not turned up any rotating dust Bianchi type III solution with the same configuration of the Killing, velocity and rotation vector fields as here. Hence, (3.16) - (3.27) is tentatively proposed as a new solution.

**Case 1.1.1.2.2:** $A_{13} = 0$.

The invariant meaning of $A_{13} = 0$ is that the Killing field $k_{(1)}^\alpha$ is everywhere orthogonal to the rotation field. We go back to (3.11). The equation $G_{13} = 0$ (in the coordinates of (3.4) with later simplifications) implies:

$$K_{22} = A_{22} K_{11}. \tag{3.28}$$

The equation $G_{22} + G_{33} = 2\Lambda$ is integrated with the result:

$$(K_{11} K_{11,z})^2 = \Lambda K_{11}{}^4 - bK_{11}{}^2 + C, \tag{3.29}$$

where $C$ is an arbitrary constant. Using this result in $G_{22} = \Lambda$ we obtain:

$$C = 1/(4A^2). \tag{3.30}$$

The case $K_{11,z} = 0$ leads to the Gödel solution [2]. When $K_{11,z} \neq 0$, the coordinate transformation:

$$K_{11}(z) = Z \tag{3.31}$$

leads to the following solution:

$$ds^2 = (dt + ydx)^2 - (Zdx)^2 - [AZ(-bydx + dy)]^2 - [\Lambda Z^2 - b^2 + 1/(2AZ)^2]^{-1} dZ^2. \tag{3.32}$$

The matter-density is:

---

[3]The method used in the search will be disclosed in Paper 3.

$$\kappa\epsilon = -3\Lambda + 1/(A^2 Z^4). \qquad (3.33)$$

With such a simple solution as (3.32) it is rather improbable that it could be still unknown. Nevertheless, the remark made after eq. (3.27) applies also here.

The subcase $\Lambda = 0$ of (3.32) is a coordinate transform of the subcase $A = 0$ of (3.16) - (3.27).

**IV. Case 1.1.2:** $b = 0$.

We go back to (2.8). From eqs. (2.3) - (2.8) we have:

$$\phi = axy + cC_3 x + \alpha(y), \qquad \lambda_2 = c\lambda_3 x + \beta(y), \qquad (4.1)$$

where $\alpha$ and $\beta$ are unknown functions and the other symbols are constants. Applying to the resulting Killing vector fields the change of basis $k'_{(2)} = k_{(2)} - (C_2/C_3)k_{(3)}$ and the coordinate transformation (1.13) with $H = \alpha/(ay + cC_3)$, $T_{,y} = (\beta - c\lambda_3 H)/(ay + cC_3)$ (the denominators are nonzero because of the assumption $\phi_{,x} \neq 0$ defining case 1), we obtain the result:

$$k^\alpha_{(2)} = cC_3 x \delta^\alpha_0 + ax \delta^\alpha_1 - (ay + cC_3)\delta^\alpha_2 + c\lambda_3 x \delta^\alpha_3, \qquad (4.2)$$

the other Killing fields being as in (2.1). The commutation relations are:

$$[k_{(1)}, k_{(2)}] = ak_{(1)} + ck_{(3)}, \qquad [k_{(1)}, k_{(3)}] = 0 = [k_{(2)}, k_{(3)}]. \qquad (4.3)$$

The Bianchi type of this algebra is III when $a \neq 0$ and II when $a = 0 \neq c$ ($a = c = 0$ is excluded by the assumption $\phi_{,x} \neq 0$ made in case 1).

In order to adapt the coordinates to $k_{(1)}$ and $k_{(3)}$, the following coordinate transformation is now carried out:

$$(t', x', y') = (t, x, y), \qquad z' = -\lambda_3 t + C_3 z, \qquad (4.4)$$

after which (primes dropped):

$$u^\alpha = \delta^\alpha_0 - \lambda_3 \delta^\alpha_3, \qquad k^\alpha_{(2)} = cx\delta^\alpha_0 + ax\delta^\alpha_1 - (ay + c)\delta^\alpha_2, \qquad (4.5)$$

but for solving the Killing equations, the cases $a \neq 0$ and $a = 0$ have to be considered separately. We first consider:

**Case 1.1.2.1:** $a \neq 0$.

Then the Killing equations imply:

$$g_{00} = 1 + \lambda_3^2 g_{33}, \qquad g_{01} = y + \lambda_3 g_{13}, \qquad g_{02} = \lambda_3 g_{23}, \qquad g_{03} = \lambda_3 g_{33},$$

$$g_{11} = -\frac{2c}{a}y - (\frac{c}{a})^2 + (\frac{c\lambda_3}{a})^2 g_{33} - \frac{2c\lambda_3}{a}(ay+c)h_{13} + (ay+c)^2 h_{11},$$

$$g_{12} = -\frac{c\lambda_3}{a(ay+c)}h_{23} + h_{12}, \qquad g_{13} = -\frac{c\lambda_3}{a}g_{33} + (ay+c)h_{13},$$

$$g_{22} = h_{22}/(ay+c)^2, \qquad g_{23} = h_{23}/(ay+c), \qquad (4.6)$$

where $g_{33}(z)$ and all $h_{ij}(z)$ are arbitrary functions. In order to introduce an orthonormal tetrad, the following new functions are defined:

$$K_{33}{}^2 = -g_{33}, \qquad H_{22} = h_{22} - h_{23}{}^2/g_{33}, \qquad K_{22}{}^2 = -H_{22},$$
$$H_{12} = (h_{12} - h_{13}h_{23}/g_{33})/H_{22},$$
$$K_{11}{}^2 = 1/a^2 + h_{13}{}^2/g_{33} + H_{12}{}^2 H_{22} - h_{11},$$
$$H_{13} = h_{13}/g_{33}, \qquad H_{23} = h_{23}/g_{33}. \tag{4.7}$$

After the coordinate transformation:

$$t = T + cx/a, \qquad y = (e^{aY} - c)/a, \qquad (x', z') = (x, z) \tag{4.8}$$

the metric form becomes:

$$ds^2 = (dT + a^{-1}e^{aY}dx)^2 - (K_{11}e^{aY}dx)^2 - [K_{22}(e^{aY}H_{12}dx + dY)]^2$$
$$- [K_{33}(\lambda_3 dT + H_{13}e^{aY}dx + H_{23}dY + dz)]^2, \tag{4.9}$$

and the Killing fields become:

$$k_{(1)}^\alpha = \delta_1^\alpha, \qquad k_{(2)}^\alpha = ax\delta_1^\alpha - \delta_2^\alpha, \qquad k_{(3)}^\alpha = \delta_0^\alpha. \tag{4.10}$$

No progress was made with the Einstein equations for (4.9). However, a large subset of solutions by Ellis [3] is contained here when $0 = \lambda_3 = H_{13} = H_{23} = H_{12}$ (the first condition means that the velocity field is a Killing field, the next two conditions mean that the Killing fields $k_{(1)}^\alpha$ and $k_{(2)}^\alpha$ are orthogonal to rotation; the invariant interpretation of the last condition is unknown). It is the case Ib of Ellis, but not in full generality. Eliis' function $t$, when expressed as a function of Ellis' $y(x^2)$, obeys:

$$2c^2(t^2)_{,yy} + K = 0, \tag{4.11}$$

where $c$ and $K$ are constants (it is Ellis' eq. (4.20), appropriately transformed). The $t^2$ implied by (4.11) is a polynomial of second degree. Only when the discriminant $\Delta$ of the polynomial is zero will Ellis' case Ib be a subcase of (4.9). The reason for this is that with $\Delta \ne 0$ the Ellis case Ib has a four-dimensional symmetry group acting multiply transitively on 3-dimensional orbits, and the group has no 3-dimensional subgroups. Only with $\Delta = 0$ a 3-dimensional subgroup exists.

**Case 1.1.2.2 : a = 0**

As stated before, the Bianchi type is now II. We go back to (4.5). With $a = 0 \ne c$, it can be assumed that $c = 1$ with no loss of generality. The velocity field and the Killing fields are:

$$u^\alpha = \delta_0^\alpha - \lambda_3 \delta_3^\alpha, \qquad k_{(1)}^\alpha = \delta_1^\alpha, \qquad k_{(2)}^\alpha = x\delta_0^\alpha - \delta_2^\alpha, \qquad k_{(3)}^\alpha = \delta_0^\alpha, \tag{4.12}$$

and the Killing equations imply:

$$g_{00} = 1 + \lambda_3{}^2 g_{33}, \qquad g_{01} = y + \lambda_3 g_{13}, \qquad g_{02} = \lambda_3 g_{23}, \qquad g_{03} = \lambda_3 g_{33},$$
$$g_{11} = (1 + \lambda_3{}^2 g_{33})y^2 + 2\lambda_3 y h_{13} + h_{11},$$
$$g_{12} = \lambda_3 y g_{23} + h_{12}, \qquad g_{13} = \lambda_3 y g_{33} + h_{13}, \tag{4.13}$$

where $g_{22}(z)$, $g_{23}(z)$, $g_{33}(z)$ and the $h_{ij}(z)$ are arbitrary functions. The new functions $K_{ii}$ and $H_{ij}$ are defined by (4.7) with $a \to \infty$ and $h_{23} = g_{23}$. The metric is then:

$$ds^2 = (dt + ydx)^2 - (K_{11}dx)^2 - (H_{12}dx + K_{22}dy)^2$$
$$-\{K_{33}[\lambda_3 dt + (\lambda_3 y + H_{13})dx + H_{23}dy + dz]\}^2. \tag{4.14}$$

Again, no progress with the Einstein equations was made at this level of generality. However, further progress was possible when:

$$\lambda_3 = 0, \tag{4.15}$$

which means, from (4.12), that the velocity field itself becomes a Killing field. Then, by a transformation of $z$ and a few redefinitions:

$$K_{33} = 1 \tag{4.16}$$

can be achieved, and this will be assumed from now on. Some of the subcases implied by the Einstein equations turn out to be empty, so they will be mentioned only briefly. The first alternative is $H_{23}$ being zero or nonzero. However, in both cases the Einstein equations together with coordinate transformations lead to the conclusion:

$$H_{13} = 0 \tag{4.17}$$

(which means $g_{\alpha\beta}k^\alpha_{(1)}w^\beta = 0$). Then the equation $G_{01} = 0$ is integrated with the result $H_{23} = A_{23}K_{11}K_{22}$ (the $A_{ij}$ being arbitrary constants). Further integration of the Einstein equations must proceed separately for $A_{23} \neq 0$ and for $A_{23} = 0$. However, the case $A_{23} \neq 0$ leads to a plain contradiction, and so the only case to consider is:

$$H_{23} = 0 = A_{23} \tag{4.18}$$

(which means $g_{\alpha\beta}k^\alpha_{(2)}w^\beta = 0$). The equation $G_{12} = 0$ is then integrated with the result:

$$H_{12,z} = H_{12}K_{22,z}/K_{22} + A_{12}K_{11}/K_{22}^2. \tag{4.19}$$

From $G_{11} - G_{22}$ we have:

$$(K_{11}K_{22,z} - K_{22}K_{11,z})_{,z} - A_{12}^2 K_{11}/K_{22}^3 = 0. \tag{4.20}$$

With the new variable $u(z)$ defined by:

$$u_{,z} = K_{11}/K_{22}^3, \tag{4.21}$$

eq. (4.20) can be integrated with the result:

$$(K_{11}/K_{22})^2 = -(A_{12}u)^2 - 2Bu + C := W(u), \tag{4.22}$$

where $B$ and $C$ are arbitrary constants. With the help of (4.22), $K_{11}$ is elliminated from the Einstein equations, and then $G_{33} = \Lambda$ remains as the only independent equation in the set. It may be written as:

$$2WK_{22,u}/K_{22} = A_{12}^2 u + B \pm (B^2 + CA_{12}^2 + 1 + 4\Lambda W K_{22}^4)^{1/2}. \tag{4.23}$$

When $\Lambda = 0$, this has simple elementary solutions, but several cases require separate treatment ($A_{12} \neq 0$ with $(B^2 + CA_{12}{}^2)$ being positive, zero or negative, $A_{12} = 0 \neq B$), so (4.23) will be left as the most compact notation. This is again tentatively proposed as a new solution (the metric is defined by (4.14) - (4.19), (4.22) and (4.23)).

When $A_{12} = 0$, these equations define a coordinate transform of Ellis' case Aii [3]. Then $H_{12} = B_{12}K_{22}$ from (4.19) and $B_{12} = 0$ results by a coordinate transformation. The invariant meaning of the condition $A_{12} = 0$ is not known to this author. With such an interpretation at hand, a new invariant definition of the Ellis case Aii would result.

**V. Case 1.2:** $C_3 = 0$.

We go back to eqs. (2.3) - (2.4) - (2.5), and immediately have to consider separately the cases $b \neq 0$ and $b = 0$. We take first:

**Case 1.2.1:** $b \neq 0$.

Equations (2.3a-c) then imply:

$$\phi = \alpha(y)e^{bx} - ay/b - C_2, \tag{5.1}$$

where $\alpha$ is an arbitrary function, $\alpha \neq 0$ because of the assumption $\phi_{,x} \neq 0$. Eq. (2.5h) implies:

$$\lambda_3 = \mu(y)e^{fx}, \tag{5.2}$$

where $\mu$ is an arbitrary function; $\mu \neq 0$ because otherwise the symmetry group becomes two-dimensional. Eq. (2.5l) then implies:

$$\mu = B\alpha^{f/b}, \qquad j = -af/b, \tag{5.3}$$

where $B$ is an arbitrary constant. For further integration, the cases $f \neq b$ and $f = b$ have to be considered separately.

**Case 1.2.1.1:** $b \neq f$.

The Killing fields that result here, after a simplification of the basis by $k'_{(2)} = k_{(2)} - (a/b)k_{(1)} - [c/(f-b)]k_{(3)}$, are:

$$k^\alpha_{(1)} = \delta^\alpha{}_1, \qquad k^\alpha_{(3)} = \alpha^{f/b}e^{fx}\delta^\alpha{}_3,$$

$$k^\alpha_{(2)} = e^{bx}[(\alpha - y\alpha_{,y})\delta^\alpha{}_0 + \alpha_{,y}\delta^\alpha{}_1 - b\alpha\delta^\alpha{}_2 + \beta\delta^\alpha{}_3] \tag{5.4}$$

(the basis change resulted in $a = c = 0$). This algebra is of Bianchi type VI$_h$ with the free parameter $(b^2 + f^2)/(b^2 - f^2)$. The Killing fields are further simplified by the coordinate transformation (1.13) with $H = b^{-1}\ln\alpha$, $T = \int (b\alpha)^{-1}\beta dy$, and by the subsequent transformation that leads out of the Plebański class:

$$t' = e^{-bx}t, \qquad x' = x, \qquad y' = bt + y, \qquad z' = e^{-fx}z. \tag{5.5}$$

The result is (primes dropped):

$$k^\alpha_{(1)} = -bt\delta^\alpha{}_0 + \delta^\alpha{}_1 - fz\delta^\alpha{}_3, \qquad k^\alpha_{(2)} = \delta^\alpha{}_0, \qquad k^\alpha_{(3)} = \delta^\alpha{}_3,$$

$$u^\alpha = e^{-bx}\delta^\alpha{}_0 + b\delta^\alpha{}_2, \qquad w^\alpha = ne^{-fx}\delta^\alpha{}_3. \tag{5.6}$$

The metric that results from the Killing equations is:

$$g_{00} = e^{2bx}(1 + b^2 h_{22}), \qquad g_{01} = e^{bx}(y - bh_{12}), \qquad g_{02} = -be^{bx}h_{22},$$

$$g_{03} = -be^{(b+f)x}h_{23}, \qquad g_{11} = h_{11}, \qquad g_{12} = h_{12}, \qquad g_{13} = e^{fx}h_{13},$$

$$g_{22} = h_{22}, \qquad g_{23} = e^{fx}h_{23}, \qquad g_{33} = e^{2fx}h_{33}, \tag{5.7}$$

where $h_{ij}(y), i,j = 1,2,3$ are arbitrary functions. No progress with the Einstein equations was made here.

**Case 1.2.1.2:** $b = f$.

The basis of the Killing fields is here:

$$k_{(1)}^\alpha = \delta^\alpha{}_1, \qquad k_{(3)}^\alpha = B\alpha(y)e^{bx}\delta^\alpha{}_3,$$

$$k_{(2)}^\alpha = e^{bx}[(\alpha - y\alpha_{,y})\delta^\alpha{}_0 + \alpha_{,y}\delta^\alpha{}_1 - b\alpha\delta^\alpha{}_2 + (cB\alpha x + \beta)\delta^\alpha{}_3] \tag{5.8}$$

(as before, $a = 0$ was achieved by a basis change). The Bianchi type is IV when $c \neq 0$ and V when $c = 0$. To this basis we apply the transformation (1.13) with $H = b^{-1}\ln\alpha$, $T = \int b^{-1}(\beta e^{-bH} - cBH)dy$, and then the subsequent transformation:

$$t' = e^{-bx}t, \qquad x' = x, \qquad y' = bt + y, \qquad z' = e^{-bx}(-ctx + z). \tag{5.9}$$

The result (i.e. the relevant vector fields and the metric that fulfills the Killing equations) is:

$$k_{(1)}^\alpha = -bt\delta^\alpha{}_0 + \delta^\alpha{}_1 - (ct + bz)\delta^\alpha{}_3, \qquad k_{(2)}^\alpha = \delta^\alpha{}_0, \qquad k_{(3)}^\alpha = \delta^\alpha{}_3,$$

$$u^\alpha = e^{-bx}\delta^\alpha{}_0 + b\delta^\alpha{}_2 - cxe^{-bx}\delta^\alpha{}_3, \qquad w^\alpha = ne^{-bx}\delta^\alpha{}_3,$$

$$g_{00} = e^{2bx}(1 + b^2 h_{22} - 2bcxh_{23} + c^2x^2 h_{33}), \qquad g_{01} = e^{bx}(y - bh_{12} + cxh_{13}),$$

$$g_{02} = e^{bx}(-bh_{22} + cxh_{23}), \qquad g_{03} = e^{2bx}(-bh_{23} + cxh_{33}),$$

$$(g_{11}, g_{12}, g_{22}) = (h_{11}, h_{12}, h_{22}),$$

$$(g_{13}, g_{23}) = e^{bx}(h_{13}, h_{23}), \qquad g_{33} = e^{2bx}h_{33}, \tag{5.10}$$

where, as before, the $h_{ij}$ are arbitrary functions of $y$. Also here, no progress was made with the Einstein equations.

**Case 1.2.2:** $b = 0$.

Eqs. (2.3a-c) and (2.3h) imply here:

$$\phi = axy + \alpha(y), \qquad \lambda_3 = \mu(y)e^{fx}, \tag{5.11}$$

where $\alpha(y)$ and $\mu(y)$ are arbitrary functions; $a \neq 0$ because of the assumption defining case 1. In further integration, the cases $f \neq 0$ and $f = 0$ have to be considered separately. However, $f \neq 0$ quickly leads to a contradiction with $a \neq 0$, so the only case to consider is:

$$f = 0. \tag{5.12}$$

Then, from (2.3d):

$$\lambda_2 = c\mu(y)x + \beta(y), \tag{5.13}$$

where $\beta$ is an arbitrary function, and from (2.3l):

$$\mu = Ay^{-j/a}, \tag{5.14}$$

where $A \neq 0$ is an arbitrary constant. The transformation (1.13) with $H = \alpha/(ay)$, $T = \int[(\beta - cAHy^{-j/a})/(ay)]dy$ leads now to $\alpha = \beta = 0$. With no loss of generality we can assume $A = 1$. The coordinates will become adapted to $k_{(1)}$ and $k_{(3)}$ after the subsequent transformation:

$$z' = y^{j/a}z \tag{5.15}$$

This leaves the metric, $u^\alpha$ and $w^\alpha$ in the Plebański form (with rescaled $n$), while the Killing fields and the commutation relations become:

$$k_{(1)}^\alpha = \delta_1^\alpha, \qquad k_{(2)}^\alpha = C_2\delta_0^\alpha + ax\delta_1^\alpha - ay\delta_2^\alpha + (cx - jz)\delta_3^\alpha, \qquad k_{(3)}^\alpha = \delta_3^\alpha,$$

$$[k_{(1)}, k_{(2)}] = ak_{(1)} + ck_{(3)}, \qquad [k_{(1)}, k_{(3)}] = 0, \qquad [k_{(2)}, k_{(3)}] = jk_{(3)}. \tag{5.16}$$

In solving the Killing equations the cases $a + j \neq 0$ and $a + j = 0$ have to be considered separately.

**Case 1.2.2.1:** $a + j \neq 0$.

The algebra (5.16) is then of Bianchi type $VI_h$ (the free parameter in the standard form of the commutation relations is $(1 - j/a)/(1 + j/a)$). The Killing equations lead to the following metric form:

$$g_{00} = 1, \qquad g_{01} = y, \qquad g_{02} = g_{03} = 0,$$
$$g_{11} = C^2 y^{-2j/a} h_{33} - 2Cy^{1-j/a} h_{13} + y^2 h_{11},$$
$$g_{12} = -Cy^{-1-j/a} h_{23} + h_{12}, \qquad g_{13} = -Cy^{-2j/a} h_{33} + y^{1-j/a} h_{13},$$
$$g_{22} = h_{22}/y^2, \qquad g_{23} = y^{-1-j/a} h_{23}, \qquad g_{33} = y^{-2j/a} h_{33}, \tag{5.17}$$

where the $h_{ij}$ are functions of the variable $u$ defined below, and $C$ is the constant:

$$C = c/(a + j), \qquad u = e^t y^{C_2/a}. \tag{5.18}$$

The limit $C_2 = 0$ of (5.17) is an allowed subcase and does not require separate treatment. With $C_2 = 0$ all $h_{ij}$ become functions of $t$ alone, and this is the first instance where a proper spatially homogeneous (necessarily tilted) Bianchi-type model appears in this scheme.

No progress in solving the Einstein equations was achieved here.

**Case 1.2.2.2:** $a + j = 0$.

The Bianchi type is IV when $c \neq 0$ and V when $c = 0$. The Killing equations lead here to the metric:

$$g_{00} = 1, \qquad g_{01} = y, \qquad g_{02} = g_{03} = 0,$$
$$g_{11} = (\frac{c}{a} y \ln y)^2 h_{33} + 2\frac{c}{a} y^2 \ln y h_{13} + y^2 h_{11},$$
$$g_{12} = \frac{c}{a} y h_{23} + h_{12}, \qquad g_{13} = \frac{c}{a} y^2 \ln y h_{33} + y^2 h_{13},$$

$$g_{22} = h_{22}/y^2, \qquad g_{23} = h_{23}, \qquad g_{33} = y^2 h_{33}, \tag{5.19}$$

where the $h_{ij}$ are arbitrary functions of the same $u = e^t y^{C_2/a}$ as in (5.18).

No progress with the Einstein equations was made here, either.

**VI. Case 2:** $\phi_{,x} = 0$.

We go back to eqs. (2.3). Now $\phi_{,y} \neq 0$ can be assumed because with $\phi_{,x} = \phi_{,y} = 0$ the Killing field $k^\alpha_{(2)}$ becomes spanned on $u^\alpha$ and $w^\alpha$, and this situation was already considered in Paper 1 [1].

If $h \neq 0$, then (2.3j) implies $\phi = -(g/h)y + A$, $A = $ const, and so $k^\alpha_{(2)} = (C_2 + A)\delta^\alpha_0 - (g/h)\delta^\alpha_1 + \lambda_2 \delta^\alpha_3$. Now $k'^\alpha_{(2)} = k^\alpha_{(2)} + (g/h)k^\alpha_{(1)}$ is spanned on $u^\alpha$ and $w^\alpha$, and so we are again in the domain of Paper 1. Hence, $h = 0$, and, from (2.3j), $g = 0$. In the same way it follows from (2.3b) and (2.3f) that:

$$a = b = d = e = g = h = 0. \tag{6.1}$$

From here on, the cases $\lambda_3 \neq 0$ and $\lambda_3 = 0$ have to be considered separately:

**Case 2.1:** $\lambda_3 \neq 0$.

Using (2.3h) with (6.1) in (2.3l) we obtain $f\phi_{,y} = j$. In the same way as above, we conclude from here that

$$f = j = 0. \tag{6.2}$$

Now from (2.3h) and (2.3d):

$$\lambda_3 = \lambda_3(y), \qquad \lambda_2 = c\lambda_3 x + \beta(y), \tag{6.3}$$

where $\lambda_3(y)$ and $\beta$ are arbitrary. The only equation that remains from the set (2.3) is now $cC_3 = 0$, and $\phi(y)$ is an arbitrary function. We first follow:

**Case 2.1.1:** $C_3 \neq 0$.

Then:

$$c = 0, \tag{6.4}$$

and the algebra becomes commutative (Bianchi type I). The transformation $z' = C_3 z/\lambda_3(y)$ gives the same result as if $\lambda_3 = C_3$, and then:

$$C_3 = \lambda_3 = 1 \tag{6.5}$$

may be assumed without loss of generality. The Killing fields are now:

$$k^\alpha_{(1)} = \delta^\alpha_1, \qquad k^\alpha_{(2)} = F\delta^\alpha_0 + \phi_{,y}\delta^\alpha_1 + \beta(y)\delta^\alpha_3, \qquad k^\alpha_{(3)} = \delta^\alpha_0 + \delta^\alpha_3, \tag{6.6}$$

where:

$$F := C_2 + \phi - y\phi_{,y}. \tag{6.7}$$

The Killing equations for $k^\alpha_{(1)}$ and $k^\alpha_{(3)}$ imply that the metric tensor is independent of $x$ and that it depends on $t$ and $z$ only through $u := (t - z)/2$. It may be assumed that $\phi_{,yy} \neq 0 \neq F - \beta$ because with $\phi_{,yy} = 0$ the Killing field $k'^\alpha_{(2)} = k^\alpha_{(2)} - \phi_{,y} k^\alpha_{(1)}$ is spanned on $u^\alpha$ and $w^\alpha$ (this is the domain of Paper 1), while $F - \beta = 0$ leads, through the Killing equations, to a singular metric ($\det(g_{\alpha\beta}) = 0$). Knowing this, we can adapt coordinates to all three Killing vectors by the following transformation:

$$t = Ft' + z', \qquad x = \phi_{,y}\, t' + x', \qquad y = y', \qquad z = \beta t' + z', \tag{6.8}$$

In the new coordinates, all the metric components depend only on $y'$. With primes dropped, the Killing, velocity and rotation fields and the metric tensor are as follows:

$$k_{(1)}^\alpha = \delta_1^\alpha, \qquad k_{(2)}^\alpha = \delta_0^\alpha, \qquad k_{(3)}^\alpha = \delta_3^\alpha, \qquad u^\alpha = (F - \beta)^{-1}(\delta_0^\alpha - \phi_{,y}\,\delta_1^\alpha - \beta\delta_3^\alpha),$$

$$w^\alpha = \{C_3 n / [\lambda_3(F - \beta)]\}(\delta_0^\alpha - \phi_{,y}\,\delta_1^\alpha - F\delta_3^\alpha),$$

$$g_{00} = F^2 - \beta^2 + 2y\phi_{,y}\,(F - \beta) + \beta^2 g_{33} + 2\phi_{,y}\,\beta g_{13} + \phi_{,y}^2\,g_{11},$$

$$g_{01} = y(F - \beta) + \phi_{,y}\,g_{11} + \beta g_{13},$$

$$g_{02} = \phi_{,y}\,g_{12} + F g_{23}, \qquad g_{03} = F - \beta + \beta g_{33} + \phi_{,y}\,g_{13}, \tag{6.9}$$

and all the $g_{ij}(y)$, $i, j = 1, 2, 3$ are arbitrary functions.

No progress with the Einstein equations was achieved in this case.

**Case 2.1.2:** $C_3 = 0$.

This means that the Killing field $k_{(3)}^\alpha$ is collinear with the rotation vector $w^\alpha$. Now we have to consider separately the cases $c \neq 0$ and $c = 0$.

**Case 2.1.2.1:** $c \neq 0$.

The Bianchi type of the algebra (2.2) is II in this case. The transformation (1.13) with $H = -\beta/(c\lambda_3)$ leads to:

$$\beta = 0. \tag{6.10}$$

It may be taken for granted that $F \neq 0$ ($F$ is still given by (6.7)) because with $F = \beta = 0$ the Killing equations imply $g_{13} = g_{23} = g_{33} = 0$ and so $g_{03} = 0$ from (6.9), i.e. $\det(g_{\alpha\beta}) = 0$. The transformation $z' = z/\lambda_3$ leads to the same result as:

$$\lambda_3 = 1 \tag{6.11}$$

(in fact, the transformation reshuffles the components $g_{2j}$, $j = 1, 2, 3$ among themselves, but formally the new metric still has the Plebański form). Hence (6.10) and (6.11) will be assumed. The subsequent transformation:

$$t = Ft', \qquad x = \phi_{,y}\,t' + x', \qquad y = y', \qquad z = \frac{1}{2}c\phi_{,y}\,t'^2 + z' \tag{6.12}$$

leads to (with primes dropped):

$$k_{(1)}^\alpha = \delta_1^\alpha, \qquad k_{(2)}^\alpha = \delta_0^\alpha + cx\delta_3^\alpha, \qquad k_{(3)}^\alpha = \delta_3^\alpha,$$

$$u^\alpha = F^{-1}(\delta_0^\alpha - \phi_{,y}\,\delta_1^\alpha - ct\phi_{,y}\,\delta_3^\alpha), \qquad w^\alpha = (n/\lambda_3)\delta_3^\alpha. \tag{6.13}$$

In the coordinates of (6.13) the Killing equations imply:

$$g_{00} = (C_2 + \phi)^2 - (y\phi_{,y}\,)^2 + \phi_{,y}^2\,h_{11},$$

$$g_{01} = yF - ct\phi_{,y} h_{13} + \phi_{,y} h_{11}, \qquad g_{02} = \phi_{,y} h_{12}, \qquad g_{03} = \phi_{,y} h_{13},$$

$$g_{11} = (ct)^2 g_{33} - 2cth_{13} + h_{11}, \qquad g_{12} = -ctg_{23} + h_{12}, \qquad g_{13} = -ctg_{33} + h_{13}, \qquad (6.14)$$

where $g_{22}, g_{23}, g_{33}$ and the $h_{ij}$ are arbitrary functions of $y$.

No progress with the Einstein equations was made.

**Case 2.1.2.2:** $c = 0$.

We go back to eqs. (6.1) - (6.3) with $c = C_3 = 0$. The algebra becomes commutative (Bianchi type I), but in contrast to Case 2.1.1 the Killing field $k^\alpha_{(3)}$ is here collinear with rotation. We have:

$$\lambda_2 = \beta(y), \qquad \lambda_3 = \lambda_3(y), \qquad \phi = \phi(y), \qquad k^\alpha_{(1)} = \delta^\alpha_1,$$

$$k^\alpha_{(2)} = F\delta^\alpha_0 + \phi_{,y} \delta^\alpha_1 + \beta\delta^\alpha_3, \qquad k^\alpha_{(3)} = \lambda_3 \delta^\alpha_3, \qquad (6.15)$$

where $F$ is given by (6.7). Also here $F \neq 0$, or else we are back in the domain of Paper 1. Eq. (6.11) applies here for the same reason as before, and the coordinates become adapted to the Killing fields after the transformation:

$$t = Ft', \qquad x = \phi_{,y} t' + x', \qquad y = y', \qquad z = \beta t' + z'. \qquad (6.16)$$

In the new coordinates (primes dropped):

$$k^\alpha_{(1)} = \delta^\alpha_1, \qquad k^\alpha_{(2)} = \delta^\alpha_0, \qquad k^\alpha_{(3)} = \delta^\alpha_3,$$

$$u^\alpha = F^{-1}(\delta^\alpha_0 - \phi_{,y} \delta^\alpha_1 - \beta\delta^\alpha_3), \qquad w^\alpha = (n/\lambda_3)\delta^\alpha_3. \qquad (6.17)$$

and the Killing equations imply:

$$g_{00} = (C_2 + \phi)^2 - (y\phi_{,y})^2 + \phi^2_{,y} g_{11} + 2\beta\phi_{,y} g_{13} + \beta^2 g_{33},$$

$$g_{01} = yF + \phi_{,y} g_{11} + \beta g_{13}, \qquad g_{02} = \phi_{,y} g_{12} + \beta g_{23}, \qquad g_{03} = \phi_{,y} g_{13} + \beta g_{33}, \qquad (6.18)$$

where all $g_{ij}, i, j = 1, 2, 3$ are arbitrary functions of $y$.

In general, no progress was made with the Einstein equations. However, a few authors have considered this case before, and a few simple exact solutions in this class are known. Therefore, we shall introduce the standard Bianchi-type coordinates in which $g_{02} = g_{12} = g_{23} = 0$, $g_{22} = -1$ in order to facilitate the comparison. The transformation to coordinates in which $g_{i2} = 0, i = 0, 1, 3$, is:

$$t = t' + F_0(y'), \qquad x = x' + F_1(y'), \qquad y = F_2(y') = y', \qquad z = z' + F_3(y'), \qquad (6.19)$$

where the functions $F_\alpha(y')$ obey:

$$\sum_{\alpha=0}^{3} g_{\alpha i} F_{\alpha,y'} = 0, \qquad i = 0, 1, 3. \qquad (6.20)$$

The set (6.20) may be solved for $F_{0,y'}, F_{1,y'}$ and $F_{3,y'}$ and then finding the $F_\alpha$ only requires calculation of a few integrals (which are seen to exist, although the integrands are at this point not known as explicit functions of $y'$). Having achieved $g_{i2} = 0$ in this way, we carry out the change of variable $y' = \int(-g_{22})^{1/2} dy$ (knowing that $g_{22} < 0$ because of the signature and $g_{22} = g_{22}(y)$ because of the symmetries) that leads to $g_{2'2'} = -1$. The components $g_{ij}$ in which $i \neq 2 \neq j$, the velocity and the rotation do not change after (6.19). In the new coordinates, dropping primes:

$$ds^2 = [(C_2 + \phi)dt + Y\mathrm{d}x]^2 - [k_{11}(\phi_{,Y} dt + dx)]^2 - dy^2$$
$$- \{k_{33}[(\beta + h_{13}\phi_{,Y})dt + h_{13}dx + dz]\}^2, \qquad (6.21)$$

where $Y$ is the $y$-coordinate of (6.18), and:

$$k_{11}{}^2 = y^2 - g_{11} + g_{13}{}^2/g_{33}, \qquad k_{33}{}^2 = -g_{33}, \qquad h_{13} = g_{13}/g_{33}. \qquad (6.22)$$

The subcase $\beta = h_{13} = 0$ of (6.21) was first considered by King [4], and it is known in the literature as "stationary cylindrically symmetric". King demonstrated that, even in this subcase, the problem is underdetermined: one function in the metric may be chosen arbitrarily (in (6.21) it is $k_{33}$). The limitation $\beta = h_{13} = 0$ resulted from the reflection symmetries assumed in Ref. 4, that in the coordinates of (6.21) correspond to $z \to -z$ and $(t, x) \to (-t, -x)$. King's metric ansatz can be also derived from the following assumptions, as follows from the present consideration:

1. The manifold has a 3-dimensional symmetry group whose algebra is of the Bianchi type I.

2. One Killing field $(k_{(3)}^\alpha)$ is collinear with rotation, the two others are linearly independent of $u^\alpha$ and $w^\alpha$.

3. The velocity vector field is spanned on $k_{(1)}^\alpha$ and $k_{(2)}^\alpha$ (hence $\beta = 0$).

4. The Killing fields $k_{(1)}^\alpha$ and $k_{(2)}^\alpha$ are both orthogonal to $k_{(3)}^\alpha$ (hence $h_{13} = 0$).

A few examples of explicit solutions of the Einstein equations are given in Ref. 4, among them the solutions of Maitra (Ref. 8, see below) and of Lanczos (Ref. 10, this one goes by the name of "Ehlers - van Stockum" in Ref. 4). Vishveshwara and Winicour [5] derived the same metric form as King and provided another explicit example of solution. The solution derived by Hoenselaers and Vishveshwara [6] that was supposed to be an example of the collection of Ref. 5, turned out to be a coordinate transform of the Gödel metric, see Ref. 7.

One more example was provided by Maitra [8]. In addition to the list above, it has the following invariant property.

5. The timelike Killing field $k_{(2)}$ has unit length so that $(C_2 + \phi)^2 - k_{11}{}^2\phi_{,Y}{}^2 = 1$.

The conditions 1 - 5 are still insufficient to reduce (6.21) to the Maitra solution; the following coordinate-dependent relations must hold in addition:

$$(C_2 + \phi)Y - k_{11}{}^2\phi_{,Y} = m, \qquad Y^2 - k_{11}{}^2 = m^2 - r^2, \qquad (6.23)$$

where $r(y)$ is a new coordinate defined by:

$$\ln(\frac{dy}{dr}) = -\frac{1}{4u^2}\left\{(1+u^2)^{1/2} - 1 + \frac{1}{8} - \frac{1}{4}\ln[\frac{1}{2}(1+u^2)^{1/2} + \frac{1}{2}]\right\}$$
$$u := 2r/a, \qquad a = \text{const}, \qquad (6.24)$$

and $m(r)$ is the function:

$$m = -\frac{1}{2}a\left\{(1+u^2)^{1/2} - 1 - \ln[\frac{1}{2}(1+u^2)^{1/2} + \frac{1}{2}]\right\}. \qquad (6.25)$$

This author was not able to interpret (6.23) in invariant terms.

The first three of the six solutions by Ozsváth [9] also belong here, and they are subcases of the class considered by King. All of Ozsváth's solutions have 4-dimensional symmetry groups whose orbits are the whole 4-dimensional manifolds. In order to place specific Ozsváth's solutions in the classification considered here, one has to identify 3-dimensional subgroups of Ozsváth's groups. Examples can be spotted by inspection in Ref. 9 in which different non-isomorphic 3-dimensional subgroups are contained in the same 4-dimensional group. Hence, the same Ozsváth's solutions should come up as limits in different classes of the present investigation. For unique and complete identification, the formulae for group generators are necessary, and these are not given for most of Ozsváth's solutions.

**Case 2.2:** $\lambda_3 = 0$.

We go back to eq. (6.1). Now necessarily $C_3 \neq 0$, so it may be assumed with no loss of generality that:

$$C_3 = 1, \tag{6.26}$$

and then eqs. (6.1) together with (2.3a), (2.3e) and (2.3i) imply:

$$c = f = j = 0, \tag{6.27}$$

and (2.3d) implies:

$$\lambda_2 = \beta(y). \tag{6.28}$$

In this case, necessarily $\beta \neq 0$ because with $\beta = 0$ the Killing equations imply that the determinant of the metric is zero. With (6.1) and (6.26) - (6.28) the whole set (2.3) is solved and the Killing fields are:

$$k_{(1)}^\alpha = \delta_1^\alpha, \qquad k_{(2)}^\alpha = F\delta_0^\alpha + \phi_{,y}\,\delta_1^\alpha + \beta\delta_3^\alpha, \qquad k_{(3)}^\alpha = \delta_0^\alpha, \tag{6.29}$$

where $F$ is still given by (6.7). It may be assumed with no loss of generality that:

$$C_2 = 0 \tag{6.30}$$

because this is equivalent to changing the basis to $k'^\alpha_{(2)} = k^\alpha_{(2)} - C_2 k^\alpha_{(3)}$. The Killing fields all commute to zero, so the Bianchi type is I once more, but this time with a still different position of the orbits with respect to the hydrodynamical vector fields; now the velocity field is the Killing field $k^\alpha_{(3)}$, i.e. it is tangent to the orbits.

In order to adapt the coordinates to the Killing fields we carry out the transformation:

$$t' = t - (F/\beta)z, \qquad x' = x - (\phi_{,y}/\beta)z, \qquad y' = y, \qquad z' = z/\beta. \tag{6.31}$$

The Killing fields $k^\alpha_{(1)}$ and $k^\alpha_{(3)}$ do not change, while $k^\alpha_{(2)}$, the rotation and the metric acquire the form:

$$k^\alpha_{(2)} = \delta_3^\alpha, \qquad w^\alpha = (n/\beta)(F\delta_0^\alpha + \phi_{,y}\,\delta_1^\alpha - \delta_3^\alpha),$$
$$g_{00} = 1, \qquad g_{01} = y, \qquad g_{02} = 0, \qquad g_{03} = \phi, \tag{6.32}$$

other components $g_{ij}$ are arbitrary functions of $y$. Similarly as it was done in (6.19) - (6.21), we can transform the metric to the Bianchi-type coordinates in which $g_{02} = g_{12} = g_{23} = 0$. Then:

$$ds^2 = (dt + ydx + \phi dz)^2 - (k_{11}dx)^2 - (k_{22}dy)^2 - [k_{33}(h_{13}dx + dz)]^2, \tag{6.33}$$

where $k_{11}$, $k_{22}$, $k_{33}$ and $h_{13}$ are functions of $y$. The transformation does not change (6.32). The Einstein equation $G_{01} = 0$ then implies:

$$h_{13}\phi_{,y} = 1 - Kk_{11}k_{22}/k_{33}, \tag{6.34}$$

where $K$ is a constant. With this, $G_{03} = 0$ implies:

$$\phi_{,y} k_{11}/(k_{22}k_{33}) - Kh_{13} = L = \text{const}, \tag{6.35}$$

and $G_{13} = 0$ implies:

$$-k_{33}{}^3 h_{13,y}/(k_{11}k_{22}) + K\phi = M = \text{const} \tag{6.36}$$

Eqs. (6.34) - (6.36) can be used to eliminate all derivatives of $\phi$ and $h_{13}$ from the Einstein tensor. This is done as follows:
1. The derivative of (6.34) is used to eliminate $h_{13}\phi_{,yy}$.
2. Eq. (6.34) is used to eliminate $(h_{13}\phi_{,y})^2$, $h_{13}\phi_{,y}$ and $h_{13}\phi_{,y}{}^2$.
3. From (6.35), $\phi_{,yy}$ is found and eliminated from the Einstein tensor completely.
4. From (6.36), $h_{13,yy}$ is found and eliminated from the Einstein tensor.
5. Eqs. (6.36) and (6.35) are used to eliminate $h_{13,y}$ and $\phi_{,y}$.

This procedure is designed so that the result of each step applies also in the limit $h_{13} = 0$. After it is completed, the Einstein tensor is diagonal.

The equation $G_{11} - G_{33} = 0$ is now:

$$\frac{1}{2}(K/k_{33})^2 + k_{22}{}^{-2}[-k_{22,y}k_{33,y}/(k_{22}k_{33}) + k_{33,yy}/k_{33} + k_{11,y}k_{22,y}/(k_{11}k_{22}) - k_{11,yy}/k_{11}]$$

$$-\frac{1}{2}[(Kh_{13} + L)/k_{11}]^2 - (K\phi - M)^2/k_{33}{}^4 = 0. \tag{6.37}$$

In this, we replace one power of $(Kh_{13} + L)$ from (6.35), one power of $(K\phi - M)$ from (6.36), multiply the result by $k_{11}k_{22}k_{33}$ and use (6.34) to eliminate $Kk_{11}k_{22}/k_{33}$ from $K^2 k_{11}k_{22}/k_{33}$. The result is then integrated, and the integral is:

$$(k_{11}k_{33,y} - k_{33}k_{11,y})/k_{22} - K\phi h_{13} - \frac{1}{2}L\phi + Mh_{13} + \frac{1}{2}Ky = A = \text{const}. \tag{6.38}$$

The equation $G_{11} - G_{22} = 0$ is:

$$-\frac{1}{2}(Kh_{13} + L)^2 + k_{22}{}^{-2}k_{33}{}^{-1}(-k_{22,y}k_{33,y}/k_{22} + k_{33,yy} - k_{11,y}k_{33,y}/k_{11}) = 0. \tag{6.39}$$

In this, we eliminate one power of $(Kh_{13} + L)$ using (6.35), then use (6.34) to eliminate $h_{13}\phi_{,y}$ from the result. The equation thus obtained is integrated to:

$$k_{33,y}/(k_{11}k_{22}) - \frac{1}{2}L\phi - \frac{1}{2}Ky + \frac{1}{2}K^2 \int (k_{11}k_{22}/k_{33})dy = B = \text{const}. \tag{6.40}$$

Now we introduce the new variable $u(y)$ and the new functions $F(y)$ and $G(y)$ by:

$$u_{,y} = k_{22}k_{33}/k_{11}, \qquad F = Kh_{13} + L, \qquad G = K\phi - M. \tag{6.41}$$

In these variables, eqs. (6.35) and (6.36) may be rewritten as:

$$G_{,u} = KF, \qquad F_{,u} = K(k_{11}{}^2/k_{33}{}^4)G. \tag{6.42}$$

These can be separated when each is differentiated by $u$:

$$G_{,uu} - [(Kk_{11})^2/k_{33}{}^4]G = 0,$$
$$F_{,uu} + (4k_{33,u}/k_{33} - 2k_{11,u}/k_{11})F_{,u} - [(Kk_{11})^2/k_{33}{}^4]F = 0. \tag{6.43}$$

Eqs. (6.43) determine $F$ and $G$ as functions of the $k_{ii}$ ($k_{22}$ is hidden in the definition of $u$). Then, (6.34) implicitly defines $k_{22}$ as a function of $k_{11}$ and $k_{33}$. Next, (6.40) defines $k_{33}$ as a function of $k_{11}$, and finally (6.38) defines $k_{11}(y)$. Hence, the set is in principle solvable. The one remaining Einstein equation is $G_{22} = \Lambda$. However, at this point it may be verified that $G_{22,y} = 0$ in virtue of the other equations, so $G_{22} = \Lambda$ merely defines $\Lambda$ in terms of the other constants. If $\Lambda = 0$, then $G_{22} = 0$ imposes an algebraic relation of the constants.

Suppose that $\phi = $ const. Then the coordinate transformation $t = t' - \phi z$ leads to $\phi = 0$ (which means $g_{\alpha\beta}k_{(2)}^\alpha k_{(3)}^\beta = 0$). Then (6.34) implies $K \neq 0$, and so (6.35) implies $h_{13} = $ const. Consequently, the transformation $z = z' - h_{13}x$ leads to $h_{13} = 0$ (i.e. $g_{\alpha\beta}k_{(1)}^\alpha k_{(2)}^\beta = 0$). With $\phi = h_{13} = 0$ the metric (6.32) - (6.33) becomes identical to the metric ansatz of Lanczos [10], and the Lanczos solution uniquely follows. In fact, with $\phi = $ const, the Killing field $k_{(2)}^\alpha$ becomes the second Killing field spanned on $u^\alpha$ and $w^\alpha$, and so we land in the domain of Paper 1. Hence, the definition of the Lanczos solution given above coincides with one of those from Paper 1.

The class defined by (6.32) - (6.43) contains the case Ciii of Ellis [3].

At this point, the whole collection of metrics considered in this paper is exhausted.

### VII. Concluding remarks.

In this paper, several classes of metrics were only derived. Investigation of their properties was postponed to separate projects; at this point it is not clear which of them would deserve further pursuit in the first turn, while investigating them all would expand the paper beyond acceptable limits.

All the cases considered here have the property that one of the Killing fields ($k_{(3)}$) is spanned on the fields of velocity $u^\alpha$ and rotation $w^\alpha$ of the source, while the other two are at every point linearly independent of $u^\alpha$ and $w^\alpha$. In every case, the explicit formulae for the Killing fields were found, and the Killing equations were solved for the components of the metric tensor. Progress with solving the Einstein equations differed from case to case. The complete listing of nonempty cases is given in Fig. 1, and the results obtained are as follows.

1. In case 1.1.1.1 (Bianchi type III) the final result is given by eq. (2.18). No exact solutions are known and none were derived here.

2. In case 1.1.1.2.1 (Bianchi type III) the final results are given by (3.16) - (3.27). A (most probably) new solution was found for the subcase $\Lambda = 0$, given by eqs. (3.16), (3.18), (3.19) and (3.24) - (3.27).

3. In case 1.1.1.2.2 (Bianchi type III) a (most probably) new solution was derived here, given by eq. (3.32).

4. In case 1.1.2.1 (Bianchi type III) the final result is given by eq. (4.9). No explicit solutions are known in general, but subcases of the case Ib of Ellis [3] belong here (see the paragraph after eq. (4.10)).

5. In case 1.1.2.2 (Bianchi type II) the final result is given by eqs. (4.14) - (4.23). No explicit solutions are known in general, but with $\lambda_3 = 0 = \Lambda$, a new exact solution was derived here, see comment after eq. (4.23). The case Aii of Ellis [3] is a subcase here.

6. In case 1.2.1.1 (Bianchi type $VI_h$) the final result is (5.6) - (5.7).

7. In case 1.2.1.2 (Bianchi types IV and V) the final result is (5.10).

8. In case 1.2.2.1 (Bianchi type $VI_h$), the final result is (5.17).

9. In case 1.2.2.2 (Bianchi types IV and V), the final result is (5.19).

10. In case 2.1.1 (Bianchi type I), the final result is (6.9).

11. In case 2.1.2.1 (Bianchi type II), the final result is (6.14). No explicit solutions are known in points 6 - 11.

12. In case 2.1.2.2 (Bianchi type I), the final result is (6.21). No explicit solutions are known in general, but a certain subcase (known as "stationary cylindrically symmetric") was considered by King [4] and Vishveshwara and Winicour [5]. Examples of explicit solutions were given in Refs. 4 and 5, and also by Maitra [8] and Ozsváth [9]; see the text after eq. (6.25).

13. In case 2.2 (Bianchi type I), the final result is given by eqs. (6.32) - (6.43). No explicit solutions are known.

The Bianchi types do not uniquely identify the various cases because in each case the orbit of the symmetry group has a different position with respect to the velocity and rotation fields. This is why the same Bianchi types occur in inequivalent cases.

The collection of results corresponding to all three Killing vectors being at every point linearly independent of velocity and rotation will be presented in Paper 3 (now in preparation). Paper 3 will also contain an overview of literature on solutions of Einstein's equations with rotating fluid source.

**Acknowledgement.** Calculations for this paper were done with use of the algebraic computer program Ortocartan [11 - 12].

**CAPTION TO THE DIAGRAM**

The classes of metrics considered in the paper. Arrows point from more general classes to subclasses. The numbers at arrows are the case-numbers used in the text. The first entry in each rectangle is the property defining the case; all the symbols are introduced in eqs. (2.1) - (2.3). The subsequent entries give the following information: 1. The Bianchi type of the corresponding algebra (2.2); 2. Information about the exact solutions (ES) known before or found in this paper; 3. The equation-numbers corresponding to the final result in the given case; 4. References to papers in which solutions of the given class were discussed.

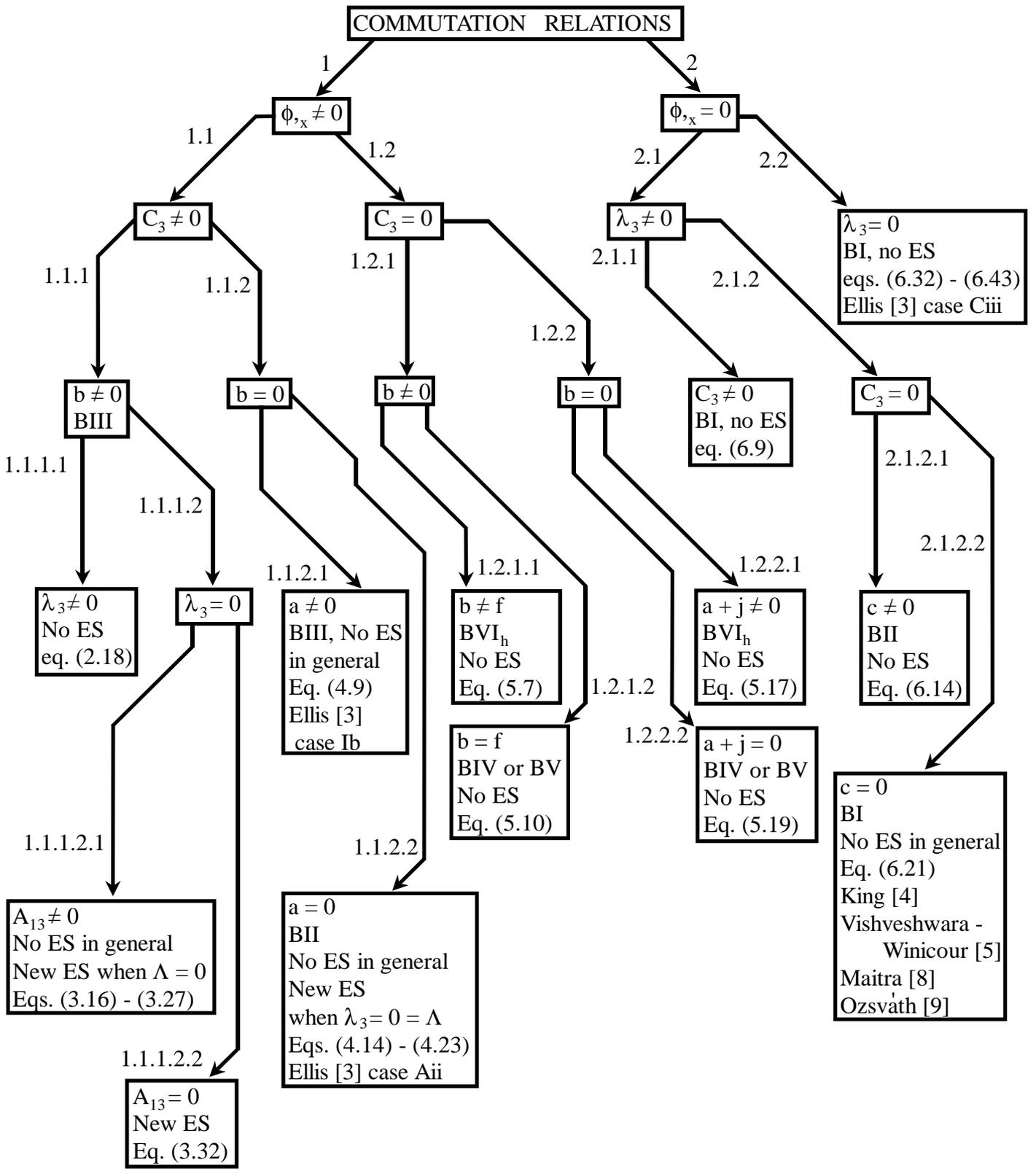